\documentclass[10pt,journal,twocolumn,final]{IEEEtran}
\usepackage{amssymb,amsmath,amsthm,amsfonts,mathtools}
\usepackage{mathrsfs}
\usepackage{graphicx}
\usepackage{subfigure}
\usepackage{color}
\usepackage{epstopdf}
\usepackage{cite}
\usepackage{algorithm}
\usepackage{algpseudocode}
\usepackage[utf8]{inputenc}
\usepackage{array}
\newtheorem{theorem}{Theorem}

\begin{document}

\title{URLLC-Aware Proactive UAV Placement in Internet of Vehicles}
\author{\IEEEauthorblockN{\begin{center}Chen-Feng Liu, \IEEEmembership{Member, IEEE}, Nirmal D. Wickramasinghe, \IEEEmembership{Student Member, IEEE},\\Himal A. Suraweera, \IEEEmembership{Senior Member, IEEE}, Mehdi Bennis, \IEEEmembership{Fellow, IEEE}, and M\'{e}rouane Debbah, \IEEEmembership{Fellow, IEEE}\end{center}}
\thanks{Chen-Feng Liu is with the Department of Informatics, New Jersey Institute of Technology, Newark, NJ 07102, USA (e-mail: chenfeng.liu@njit.edu).}
\thanks{Nirmal D. Wickramasinghe is with the Department of Electronic Engineering, Maynooth University, Maynooth, W23 A3HY Ireland (e-mail: nirmal.wickramasinghe.2023@mumail.ie).}
\thanks{Himal A. Suraweera is with the Department of Electrical and Electronic Engineering, University of Peradeniya, Peradeniya 20400, Sri Lanka (e-mail: himal@eng.pdn.ac.lk).}
\thanks{Mehdi Bennis is with the Centre for Wireless Communications, University of Oulu, 90014 Oulu, Finland (e-mail: mehdi.bennis@oulu.fi).}
\thanks{M\'{e}rouane Debbah is with the KU 6G Research Center, Khalifa University of Science and Technology, P.O. Box 127788, Abu Dhabi, United Arab Emirates, and also with CentraleSup\'{e}lec, Universit\'{e} Paris-Saclay, 91192 Gif-sur-Yvette, France  (e-mail: merouane.debbah@ku.ac.ae).}}

\maketitle

\begin{abstract}
Unmanned aerial vehicles (UAVs) are envisioned to provide diverse services from the air. The service quality may rely on the wireless performance which is affected by the UAV's position. In this paper, we focus on the UAV placement problem in the Internet of Vehicles, where the UAV is deployed to monitor the road traffic and sends the monitored videos to vehicles. The studied problem is formulated as video resolution maximization by optimizing over the UAV's position. Moreover, we take into account the maximal transmission delay and impose a probabilistic constraint. To solve the formulated problem, we first leverage the techniques in extreme value theory (EVT) and Gaussian process regression (GPR) to characterize the influence of the UAV's position on the delay performance. Based on this characterization, we subsequently propose a proactive resolution selection and UAV placement approach, which adaptively places the UAV according to the geographic distribution of vehicles. Numerical results justify the joint usage of EVT and GPR for maximal delay characterization. Through investigating the maximal transmission delay, the proposed approach nearly achieves the optimal performance when vehicles are evenly distributed, and reduces 10\% and 19\% of the 999-th 1000-quantile over two baselines when vehicles are biased distributed. 
\end{abstract}

\begin{IEEEkeywords}
Beyond 5G, Internet of Vehicles (IoV), UAV, URLLC, extreme value theory, Gaussian process regression.
\end{IEEEkeywords}

\section{Introduction}\label{Sec: Intro}
\IEEEPARstart{T}{he} Internet of Vehicles (IoV), evolved from vehicular networks  by entangling intelligence and telematics, is a key enabler for autonomous vehicles and intelligent transportation systems (ITSs) \cite{XuZhoCheLyuShiCheShe18}. 
By leveraging the advantages of agility and flexibility, unmanned aerial vehicles (UAVs) have been considered for network coverage and capacity enhancement in IoV, e.g., used as the aerial base stations or relays \cite{ShiZhoLiXuZhaShe18}, and other applications such as real-time vehicle tracking \cite{LiuZhuDenGuaWanLuoLinZha19}.
Due to the high mobility of vehicles and the distinctive feature of air-to-ground communication channels, the one-time placement and trajectory (i.e., positions in consecutive time units) design of UAVs play an important role in IoV systems \cite{ShiZhoLiXuZhaShe18}.

\subsection{Related Work}
Various wireless performance metrics such as throughput, latency, energy or power consumption, and localization accuracy were optimized in the UAV deployment problems of IoV networks \cite{LiaMaHuaWan23,LiuLaiLinLeu22,CaiFenHeXuZhaXie20,DemTokEki20,YanLiaZha21}.
While placing UAVs, the authors in \cite{LiaMaHuaWan23} took into account the signal occlusion effect in the three-dimensional (3D) highway interchanges instead of the common two-dimensional (2D) roads. Their objective was to maximize the net throughput subject to the energy budget for flight and hovering. 
Liu {\it et al.}~took into account the collision and communication interference between UAVs while maximizing the net throughput \cite{LiuLaiLinLeu22}. The authors further proposed a joint vehicle scheduling and power allocation approach along with trajectory design.
The collaborative trajectory optimization of multiple UAVs was considered in \cite{CaiFenHeXuZhaXie20}. Therein, subject to the end-to-end latency constraint, the number of UAVs which can cover the entire vehicular network was minimized.
The work \cite{DemTokEki20} minimized the total power consumption for hovering and communication. By taking into account the wireless backhaul capacity, the constraints on the end-to-end delay violation probability and downlink rate were imposed. Further, a UAV placement and transmit power allocation approach was proposed.
Yang {\it et al.}~utilized UAVs to assist vehicle localization in which the placement of UAVs will affect the localization accuracy \cite{YanLiaZha21}. To minimize a lower bound on the localization error, a joint transmit power and bandwidth allocation approach along with a UAV placement method were proposed.

\subsection{Our Contribution}
Among the existing UAV deployment works in IoV, little attention has been devoted to the reliability concern with respect to the maximal delay. Since road safety enhancement is one of the objectives of ITSs \cite{XuZhoCheLyuShiCheShe18}, ultra-reliable low-latency communication (URLLC) is of paramount importance in IoV \cite{LiuBen18,BatLiuBenSurHon20}.
We further argue that the maximal delay (in the spatial or temporal perspective) deserves the dedicated focus in IoV networks because car accidents most likely happen in the worst-cast situations.
The maximal queuing delay among all vehicles (i.e., spatial) has been considered in our previous IoV work \cite{LiuBen18}.
Hence, motivated by this shortage, we focus on the maximal transmission delay over the upcoming time period (i.e., temporal) instead of the delay in one single transmission, and further impose a probabilistic constraint on the delay threshold violation as the URLLC requirement. A scenario in which the UAV is deployed to monitor the road traffic and sends the monitored videos to vehicles is considered in this work. The higher resolution video contains the preciser traffic information but deteriorates the transmission delay performance. Therefore,  focusing on one-time UAV placement, we aim to maximize video resolution subject to the imposed URLLC constraint.
By leveraging the results in extreme value theory (EVT), the maximal transmission delay and the URLLC constraint are first characterized by the generalized extreme value (GEV) distribution. Then given the historical data of the maximal delay, an offline approach for learning the mathematical expression of the impact of the UAV's position on the GEV distribution is proposed with the aid of Gaussian process regression (GPR). Subsequently, by solving the resolution maximization problem along with the learned results, we propose an approach which can proactively place the UAV and select the optimal resolution when the network starts operating, i.e., in an online manner.
In numerical results, by showing the coincide of the empirical distribution and the approximate GEV distribution (estimated by GPR) of the maximal delay, the effectiveness of utilizing EVT and GPR for maximal delay characterization is verified. When vehicles are evenly distributed, the proposed UAV placement approach nearly achieves the optimal delay performance. Compared with two baselines which, respectively, fixes the UAV's position and randomly deploys the UAV, our approach attains 10\% and 19\% reductions, with respect to the 999-th 1000-quantile of the maximal transmission delay, in the biased vehicle distribution. Furthermore, by showing that the UAV can be placed around the vehicle group center as the vehicle distribution varies, we demonstrate the adaptability of our proposed approach for UAV placement.

\section{System Model and Problem Formulation}\label{Sec: System}

\subsection{System Architecture}
\begin{figure}[t]
\centering
\includegraphics[width=\columnwidth]{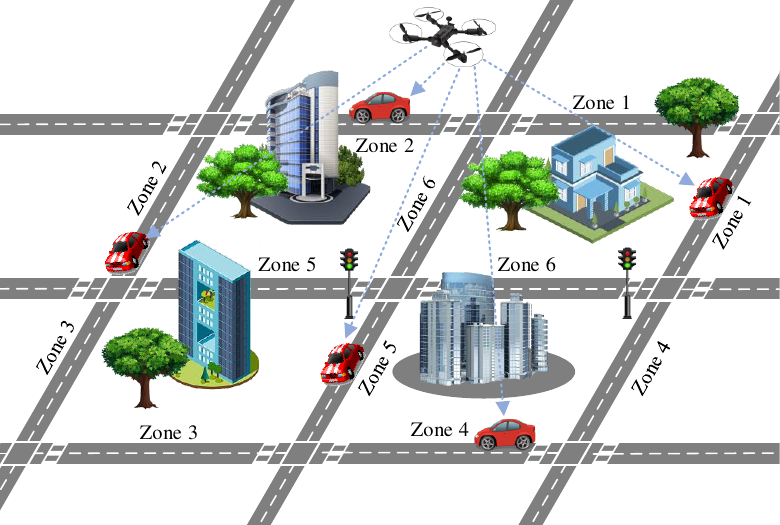}
\caption{Studied IoV network with a UAV in which the longitudinal and latitudinal roads are divided into six zones.}
\label{Fig: System}
\end{figure}

As shown in Fig.~\ref{Fig: System}, we consider an IoV system consisting of a UAV and a set $\mathcal{V}$ of $V$ vehicular user equipments (VUEs) in a geographic area. The UAV is deployed to continuously monitor and record the road traffic of the whole area. The VUE can extract information from the recorded traffic video for further usage. To this goal, the UAV sends the recorded videos to the VUEs. Moreover, the UAV records the videos with $L$ different resolution levels and then sends separate videos (with different resolutions) to different VUEs.
The lower the resolution level is, the worse the video quality which consequently affects the accuracy of the extracted traffic information \cite{YuTakKaiSak21}.

Regarding the transmission timeline, we focus on a $S_0$-slot time block $\mathcal{S}=\{1,\cdots,S_0\}$, in which the time slot is indexed by $\tau\in\mathcal{S}$. Since a video is composed of the image sequence, we simply assume that for a single VUE, one image is sent in each time slot, and then use the terminology ``image'' instead of ``video'' hereafter. In addition, the UAV equally and orthogonally allocates its dedicated bandwidth $W$ and total transmit power $P$ to all VUEs. Provided an image size $A_{l_i}$, the transmission delay from the UAV to VUE $i\in\mathcal{V}$ in time slot $\tau\in\mathcal{S}$ is given by $T_{i}(\tau)= \frac{A_{l_i}V }{W\log_2\big(1+\frac{Ph_{i}(\tau)}{N_0W}\big)}$, in which $N_0$ and $h_{i}(\tau)$ denote the power spectral density of the additive white Gaussian noise and the channel gain between the UAV and VUE $i$ in slot $\tau$, respectively. 
Referring to the practical video specification, we consider the discrete resolution level $l_i\in\mathcal{L}=\{1,\cdots,L\}$ for each VUE $i$ with the corresponding image size $A_{l_i}$.
The resolution levels of the images for all VUEs are decided at the beginning of  $\mathcal{S}$ and fixed during the entire time block. Moreover, the VUEs are moving during $\mathcal{S}$, whereas the UAV stays static at a fixed position.

 \begin{figure*}[t]
\centering
\subfigure[$\hat{\mu}$ versus $\mathbf{e}_{\rm u}.$]{\includegraphics[width=0.67\columnwidth]{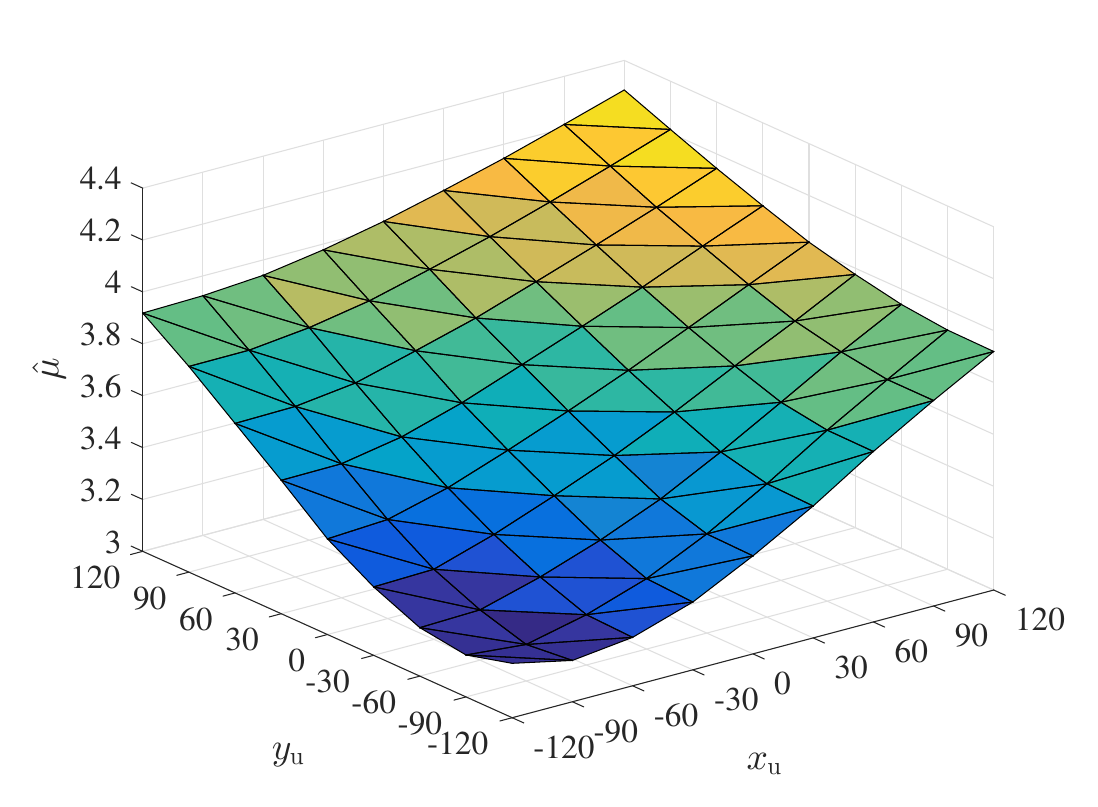}}
\subfigure[$\hat{\sigma}$ versus $\mathbf{e}_{\rm u}$.]{\includegraphics[width=0.67\columnwidth]{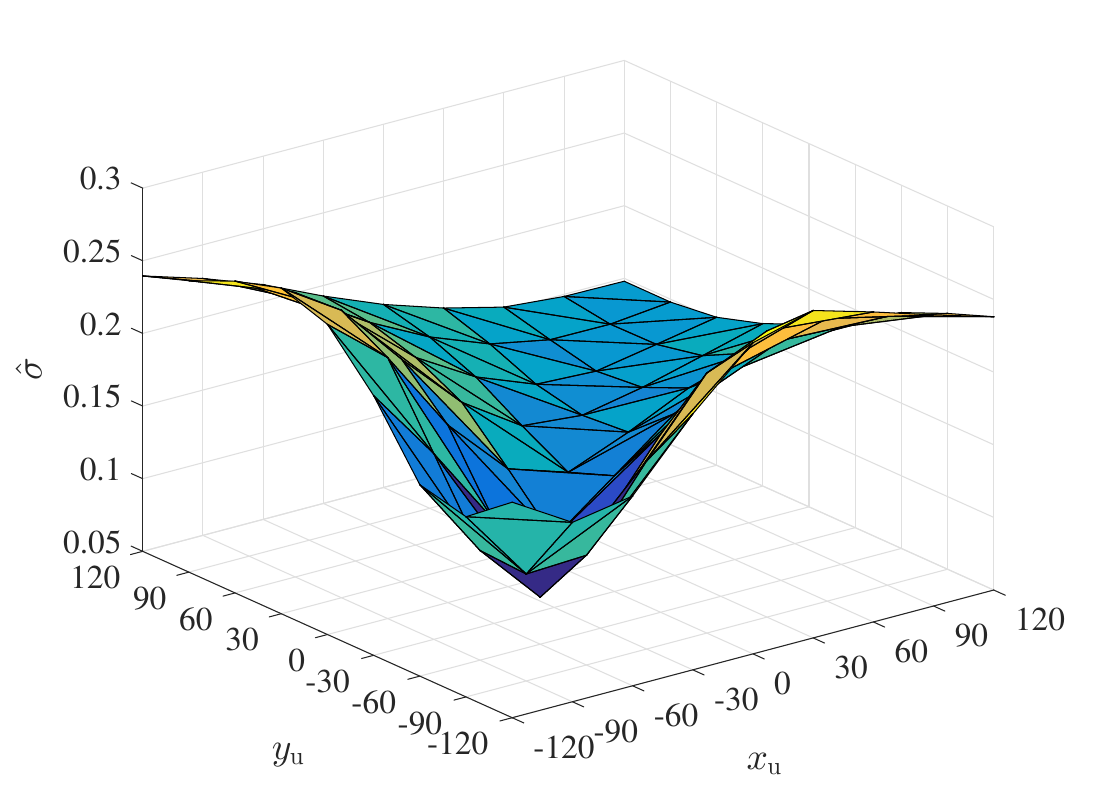}}
\subfigure[$\hat{\xi}$ versus $\mathbf{e}_{\rm u}$.]{\includegraphics[width=0.67\columnwidth]{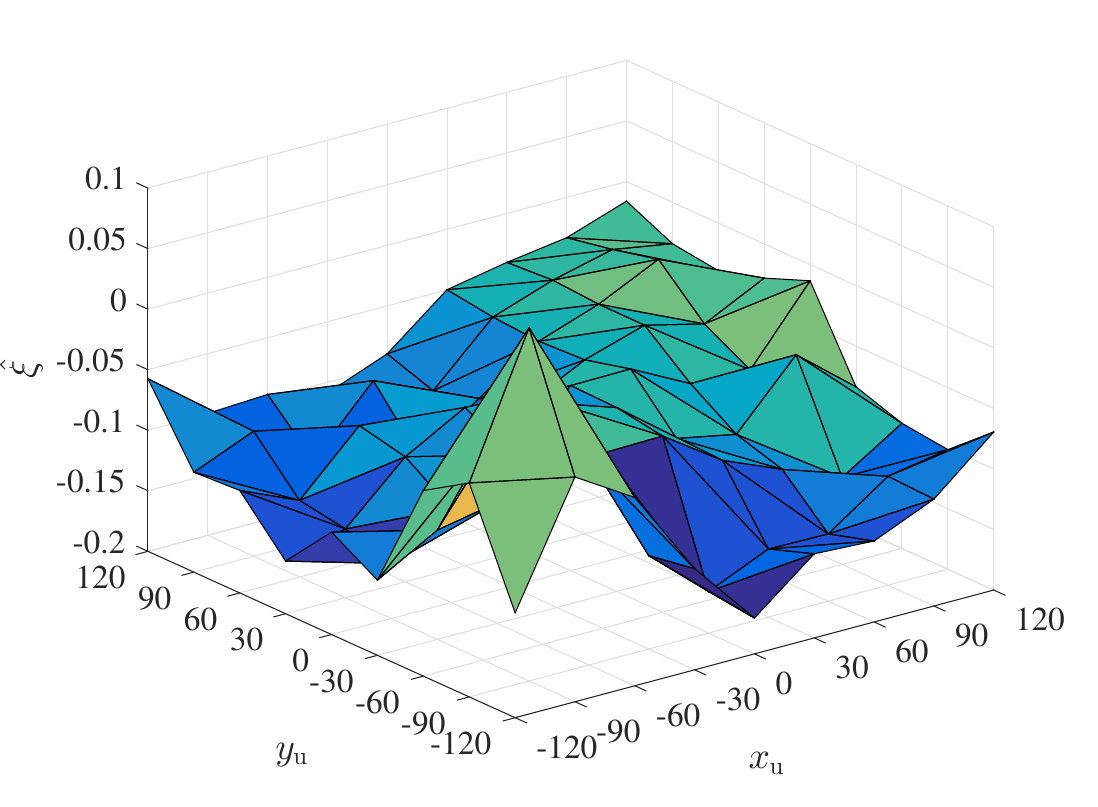}}
\caption{Learned GEV distribution parameters $\hat{\mu}$, $\hat{\sigma}$, and $\hat{\xi}$ (in Zone 3) versus $\mathbf{e}_{\rm u}$ given that the estimation errors are not significant.}
	\label{Fig: GPR example}
\end{figure*}

\subsection{Problem Formulation}
As motivated in Section \ref{Sec: Intro}, we are concerned about the VUE's experienced maximal transmission delay over the time block $\mathcal{S}$, which is defined as $T_i^{(\max)}=\underset{\tau\in\mathcal{S}}{\max}\{T_{i}(\tau)\}$ for each VUE $i$. Given the uncertain movement of the VUE and the wireless channel randomness, the maximal transmission delay $T_i^{(\max)}$ is stochastic. Taking into account these uncertainties, we impose a probabilistic constraint $\Pr\{T_i^{(\max)}> T_{i}^ {\rm th}\}\leq \epsilon$ on each VUE $i$'s maximal transmission delay as the URLLC requirement. Here, $\Pr\{\cdot\}$, $T_{i}^{\rm th}$, and $\epsilon\ll 1$ are the probability measure, the delay threshold, and the tolerable probability of delay violation, respectively. Note that besides the aforementioned uncertainties, the transmission delay also depends on the distance between the UAV and VUE (reflected by the channel gain) and the resolution level (i.e., image size). Regarding the UAV-to-VUE distance, in this work we aim to place the UAV at a position, denoted by the 3D coordinate  $\mathbf{e}_{\rm u}=(x_{\rm u},y_{\rm u},H)\in\mathbb{R}^{3}$, such that the URLLC requirements for all VUEs are satisfied. On one hand, the lower transmission delay and delay violation probability can be achieved by the lower resolution level. On the other hand, the higher resolution level is preferable for better information extraction. Accordingly, our studied optimization problem is cast as
\begin{subequations}\label{Eq: problem}
\begin{IEEEeqnarray}{cl}
\underset{\mathbf{e}_{\rm u},\mathbf{L}}{\mbox{maximize}}&~~\sum\limits_{i\in\mathcal{V}}A_{l_i}\label{Eq: problem 1}
\\\mbox{subject to}&~~\Pr\{T_i^{(\max)}> T_{i}^ {\rm th}\}\leq \epsilon,~\forall\,i\in\mathcal{V},\label{Eq: problem 2}
\\&~~(x_{\rm u}-x_{\rm 0})^2+(y_{\rm u}-y_{\rm 0})^2\leq d_{\rm th}^2,\label{Eq: problem 3}
\\&~~l_i\in\mathcal{L},~\forall\,i\in\mathcal{V},\label{Eq: problem 4}
\end{IEEEeqnarray}
\end{subequations}
in which the objective is to maximize the sum resolution over all VUEs' images. For notational simplicity, $\mathbf{L}=(l_{i}:i\in\mathcal{V})$ denotes the network-wide resolution level vector. Assuming that the UAV stays at the same altitude $H>0$ in this work, we move the UAV from the original position $(x_{\rm 0},y_{\rm 0},H)$ to a new position $(x_{\rm u},y_{\rm u},H)$ with the maximal displacement $d_{\rm th}$ in constraint \eqref{Eq: problem 3}. Hence, for simplicity, the 2D coordinate $(x_{\rm u},y_{\rm u})$ for $\mathbf{e}_{\rm u}$ is considered hereafter. Note that problem \eqref{Eq: problem} is solved before the transmission timeline $\mathcal{S}$ starts.

\section{Maximal Delay Characterization}
\label{Sec: Delay characterization}

\subsection{Generalized Extreme Value Distribution}

To solve problem \eqref{Eq: problem}, we require the complementary cumulative distribution function (CCDF) of the maximal transmission delay $T_i^{(\max)}$ in closed-form. However, since it is cumbersome to have the statistics of the VUE's movement, the CCDF of $T_i^{(\max)}$ is not available. To address this dilemma, let us first introduce a fundamental theorem in EVT \cite[p.~96]{Coles01}.
\begin{theorem}\label{Thm: GEV}
Let $T(1),T(2),\cdots$ be a stationary process with the identical marginal distribution and define $T_{\max}=\max\{T(1),\cdots,T(k)\}$. As $k\to\infty$, $T_{\max}$ can be approximately characterized by a GEV distribution $G(t;\mu,\sigma,\xi)$, i.e., 
$\Pr\{T_{\max}> t\}\approx G(t;\mu,\sigma,\xi)\equiv 1- e^{-(1+\xi(t-\mu)/\sigma)^{-1/\xi}}$,
with a location parameter $\mu\in\mathbb{R}$, a scale parameter $\sigma>0$, and a shape parameter $\xi\in\mathbb{R}$.
\end{theorem}
The VUE's locations in the consecutive time slots are neighbored, correlating the transmission delays. At the beginning of $\mathcal{S}$, VUE $i$ may locate at any position of the considered geographic area. Thus, the transmission delay process $\{T_i(\tau):\tau\in\mathcal{S}\},\forall\,i\in\mathcal{V},$ is stationary with the same marginal distribution. Since frequently changing video resolution is not practical, we can assume a sufficiently large $S_0$ in the considered system. Then by applying the results in Theorem \ref{Thm: GEV} and $\ln(1-\epsilon)\approx -\epsilon$ to \eqref{Eq: problem 2}, we can approximate problem \eqref{Eq: problem} as
\begin{subequations}\label{Eq: problem EVT}
\begin{IEEEeqnarray}{cl}
\hspace{-1em}\underset{\mathbf{e}_{\rm u},\mathbf{L}}{\mbox{maximize}}&~~\sum\limits_{i\in\mathcal{V}}A_{l_i}
\\\hspace{-1em}\mbox{subject to}&~~(1+\xi_i(T_{i}^{\rm th}-\mu_i)/\sigma_i)^{-1/\xi_i}\leq \epsilon,~\forall\,i\in\mathcal{V},\label{Eq: EVT constraint}
\\\hspace{-1em}&~~\eqref{Eq: problem 3} \mbox{ and }\eqref{Eq: problem 4},\notag
\end{IEEEeqnarray}
\end{subequations}
where $(\mu_i,\sigma_i,\xi_i)$ are the characteristic parameters of the approximate GEV distribution of $T_i^{(\max)}$. As highlighted in Section \ref{Sec: System}, the transmission delay in each slot $T_i(\tau)$ and maximal transmission delay $T_i^{(\max)}$ are affected by the UAV-to-VUE distance and image size. Thus, the parameters $(\mu_i,\sigma_i,\xi_i)$ in \eqref{Eq: EVT constraint} are functions of $\mathbf{e}_{\rm u}$ and $l_i$. In other words, to proceed with problem \eqref{Eq: problem EVT}, we need to have the mapping $(\mathbf{e}_{\rm u},l_i)\rightarrow(\mu_i,\sigma_i,\xi_i)$.
To this end, let us define $\tilde{T}_{i}^{(\max)}=\underset{\tau\in\mathcal{S}}{\max}\Big\{ \frac{A_{1}}{W\log_2\big(1+\frac{Ph_{i}(\tau)}{N_0W}\big)}\Big\}$, which can also be characterized by a GEV distribution $G(t;\tilde{\mu}_i,\tilde{\sigma}_i,\tilde{\xi}_i)$. 
Then by rewriting VUE $i$'s maximal transmission delay as $T_{i}^{(\max)}=\tilde{T}_{i}^{(\max)}V A_{l_i}/A_1$ and incorporating it with $G(t;\mu_i,\sigma_i,\xi_i)$ and $G(t;\tilde{\mu}_i,\tilde{\sigma}_i,\tilde{\xi}_i)$, we can derive
\begin{subequations}\label{Eq: single GEV}
\begin{align}
\mu_i&=
\tilde{\mu}_iV A_{l_i}/A_1,
\\\sigma_i&=
\tilde{\sigma}_i V A_{l_i}/A_1,
\\\xi_i&=
\tilde{\xi}_i,
\end{align}
\end{subequations}
which explicitly show the impact of $l_i$ on $(\mu_i,\sigma_i,\xi_i)$. Next we elaborate the methods to learn the characteristic parameters $(\tilde{\mu}_i,\tilde{\sigma}_i,\tilde{\xi}_i)$ and find how the UAV's position $\mathbf{e}_{\rm u}$ influences these parameter values.

\subsection{GEV Distribution Parameters Characterization}\label{Sec: GEV characterization}

Assuming that there are abundant historical data samples of the maximal delay $\tilde{T}_{i}^{(\max)}$, we learn the characteristic parameters $(\tilde{\mu}_i,\tilde{\sigma}_i,\tilde{\xi}_i)$ from these independent and identically distributed data. Specifically, the geographic area is divided into multiple disjoint zones $\mathcal{Z}$ as shown in Fig.~\ref{Fig: System}. 
While collecting the historical data, if the VUE locates in Zone $z\in\mathcal{Z}$, we attribute its experienced maximal delay $\tilde{T}_i^{(\max)}$ to the set $\mathcal{T}_z=\{t^1_z,\cdots,t^n_z,\cdots,t^N_z\}$.
During the $S_0$ slots to measure a maximal delay sample $t^n_z$, the UAV stays static in a fixed position. Additionally, the UAV's position is identical for all data samples in $\mathcal{T}_z$.
Having the sufficient amount of data samples, we subsequently aim to find the GEV distribution $G(t;\tilde{\mu}_z,\tilde{\sigma}_z,\tilde{\xi}_z)$, which is the closest to the empirical distribution of $\mathcal{T}_z$ with respect to the Kullback--Leibler divergence. We further note that by straightforward derivations, the closest GEV distribution can be obtained via maximum likelihood estimation (MLE)
$\underset{\tilde{\mu}_z,\tilde{\sigma}_z,\tilde{\xi}_z}{\arg\max}~\frac{1}{N}\sum_{n=1}^{N}\ln\big(f(\tilde{\mu}_z,\tilde{\sigma}_z,\tilde{\xi}_z|t_z^n)\big)$,
where  $f(\tilde{\mu}_z,\tilde{\sigma}_z,\tilde{\xi}_z|t_z^n)= \frac{1}{\tilde{\sigma}_z} e^{-(1+\tilde{\xi}_z(t^n_z-\tilde{\mu}_z)/\tilde{\sigma}_z)^{-1/\tilde{\xi}_z}}
\cdot (1+\tilde{\xi}_z(t^n_z-\tilde{\mu}_z)/\tilde{\sigma}_z)^{-1/\tilde{\xi}_z-1}$ is the likelihood function of the GEV distribution. Since deriving the closed-form solution to MLE is not feasible, we resort to the gradient ascent method and iteratively update (till achieving convergence)
\begin{multline}\label{Eq: GEV learning}
[\tilde{\mu}_z^{j+1}, \tilde{\sigma}_z^{j+1},\tilde{\xi}_z^{j+1}]= [\tilde{\mu}_z^{j}, \tilde{\sigma}_z^{j},\tilde{\xi}_z^{j}]
\\+\frac{\nu}{N}\sum_{n=1}^{N}\nabla\ln\big(f(\tilde{\mu}_z^{j}, \tilde{\sigma}_z^{j},\tilde{\xi}_z^{j}|t_z^n)\big).
\end{multline}
Here, $\nu$ is the learning rate while $j$ denotes the iteration index.
The details of the gradient in \eqref{Eq: GEV learning} are given as follows:
\begin{equation*}
\begin{cases}
\frac{\partial}{\partial \tilde{\mu}}\ln\big(f(\tilde{\mu}_z^{j}, \tilde{\sigma}_z^{j},\tilde{\xi}_z^{j}|t_z^n)\big)  
=\frac{\tilde{\xi}_z^j+1}{\tilde{\sigma}_z^j+\tilde{\xi}_z^j(t^n_z-\tilde{\mu}_z^j)} 
\\\hspace{9.6em}-\frac{1}{\tilde{\sigma}_z^j}\big(1+\tilde{\xi}_z^j(t^n_z-\tilde{\mu}_z^j)/\tilde{\sigma}_z^j\big)^{-1/\tilde{\xi}_z^j-1},
\\\frac{\partial}{\partial \tilde{\sigma}}\ln\big(f(\tilde{\mu}_z^{j}, \tilde{\sigma}_z^{j},\tilde{\xi}_z^{j}|t_z^n)\big)  
= \frac{ (\tilde{\xi}_z^j+1)(t^n_z-\tilde{\mu}_z^j)}{ (\tilde{\sigma}_z^j)^2+\tilde{\xi}_z^j(t^n_z-\tilde{\mu}_z^j)\tilde{\sigma}_z^j}
\\\hspace{5.2em} -\frac{(t^n_z-\tilde{\mu}_z^j)}{(\tilde{\sigma}_z^j)^2}\big(1+\tilde{\xi}_z^j(t^n_z-\tilde{\mu}_z^j)/\tilde{\sigma}_z^j\big)^{-1/\tilde{\xi}_z^j-1} -\frac{1}{\tilde{\sigma}_z^j},
\\\frac{\partial}{\partial \tilde{\xi}}\ln\big(f(\tilde{\mu}_z^{j}, \tilde{\sigma}_z^{j},\tilde{\xi}_z^{j}|t_z^n)\big)  
=\frac{1}{(\tilde{\xi}_z^j)^2}\ln \big(1+\tilde{\xi}_z^j(t^n_z-\tilde{\mu}_z^j)/\tilde{\sigma}_z^j\big)
\\\hspace{3em}-\frac{(1/\tilde{\xi}_z^j+1)(t^n_z-\tilde{\mu}_z^j)}{ \tilde{\sigma}_z^j+\tilde{\xi}_z^j(t^n_z-\tilde{\mu}_z^j)}
 -\frac{ (1+\tilde{\xi}_z^j(t^n_z-\tilde{\mu}_z^j)/\tilde{\sigma}_z^j)^{-1/\tilde{\xi}_z^j}}{(\tilde{\xi}_z^j)^2}
\\\hspace{5.3em}\times \Big[\ln\big(1+\tilde{\xi}_z^j(t^n_z-\tilde{\mu}_z^j)/\tilde{\sigma}_z^j\big)- \frac{\tilde{\xi}_z^j (t^n_z-\tilde{\mu}_z^j)}{\tilde{\sigma}_z^j+\tilde{\xi}_z^j(t^n_z-\tilde{\mu}_z^j)}\Big].
\end{cases}
\end{equation*}
If VUE $i$ locates in Zone $z$, we consider $(\tilde{\mu}_z,\tilde{\sigma}_z,\tilde{\xi}_z)$ as the GEV distribution parameters of VUE $i$'s experienced maximal delay in \eqref{Eq: single GEV}, i.e.,   $(\tilde{\mu}_i,\tilde{\sigma}_i,\tilde{\xi}_i)=(\tilde{\mu}_z,\tilde{\sigma}_z,\tilde{\xi}_z)$.

\subsection{GPR for Position-Parameter Mapping}

When we learn $(\tilde{\mu}_z,\tilde{\sigma}_z,\tilde{\xi}_z)$ for each Zone $z$ in \eqref{Eq: GEV learning}, a specific position $\mathbf{e}_{\rm u}$ of the UAV is considered. As emphasized previously, we require the mapping function $\mathbf{e}_{\rm u}\rightarrow(\tilde{\mu}_z,\tilde{\sigma}_z,\tilde{\xi}_z),\forall\,\mathbf{e}_{\rm u}$. That is, the corresponding GEV distribution parameters of the continuous UAV positions. However, learning the parameters $(\tilde{\mu}_z,\tilde{\sigma}_z,\tilde{\xi}_z)$ with the infinite or extremely large amount of positions is impractical and infeasible. Alternatively, by resorting to the regression concept, we consider the moderate number of positions and use them with the corresponding learned parameters to approximate the mapping function. 
Note that the mathematical model of the mapping $\mathbf{e}_{\rm u}\rightarrow(\tilde{\mu}_z,\tilde{\sigma}_z,\tilde{\xi}_z)$ is unknown. Directly using the simple regression techniques such as linear regression, polynomial regression, or the alike with other non-linear models (e.g., logarithm) may not be suitable. This impropriety can be understood from Fig.~\ref{Fig: GPR example}, where a compact mathematical model for $\mathbf{e}_{\rm u}\rightarrow \tilde{\xi}_z$ cannot be straightforwardly found. To circumvent this issue, while artificial neural networks can well approximate the non-linear and complex functions without the prior model knowledge \cite{AlsLinNiyTan14}, GPR provides a clean-slate mathematical expression (which, in general, is not available in neural network-based methods) for the mapping function $\mathbf{e}_{\rm u}\rightarrow(\tilde{\mu}_z,\tilde{\sigma}_z,\tilde{\xi}_z)$ in a statistical manner based on the learned parameters, i.e., the training data \cite{RasWil06}. In contrast with artificial neural networks, less training data is required in GPR \cite{AlsLinNiyTan14}. Therefore, we leverage GPR among the regression techniques.

Before applying GPR, we introduce an auxiliary parameter $\tilde{\zeta}$ (for the ease of selecting the resolution level in \eqref{Eq: Selection and positioning problem} afterwards) through a one-to-one function $\tilde{\zeta}=f(\tilde{\xi})= \frac{\epsilon^{-\tilde{\xi}}-1}{\tilde{\xi}}>0$ given $0<\epsilon<1$. Hereafter, we consider $\tilde{\zeta}$ instead of the shape parameter $\tilde{\xi}$.
For each Zone $z$, we randomly select $K$ positions of the UAV and learn the corresponding GEV distribution parameters, which are collectively denoted by a set $\mathcal{D}_z=\{(\bar{\mathbf{e}}_{\rm u}^k;\hat{\mu}_z^k,\hat{\sigma}_z^k,\hat{\zeta}_z^k):1\leq k\leq K\}$. By applying the data samples in $\mathcal{D}_z$ to GPR, the values of the parameters $\tilde{\mu}_z$, $\tilde{\sigma}_z$, and $\tilde{\zeta}_z$ at any arbitrary position $\mathbf{e}_{\rm u}$ are estimated  as 
\begin{equation}\label{Eq: Regression}
\begin{cases}
\tilde{\mu}_z(\mathbf{e}_{\rm u})=[c_{\mu_z}^1,\cdots, c_{\mu_z}^K](\mathbf{C}_{\mu_z}+\lambda_{\mu_z}\mathbf{I}_K)^{-1}[\hat{\mu}_z^1,\cdots, \hat{\mu}_z^K]^{\rm T},
\\\tilde{\sigma}_z(\mathbf{e}_{\rm u})=[c_{\sigma_z}^1,\cdots, c_{\sigma_z}^K](\mathbf{C}_{\sigma_z}+\lambda_{\sigma_z}\mathbf{I}_K)^{-1}[\hat{\sigma}_z^1,\cdots, \hat{\sigma}_z^K]^{\rm T},
\\\tilde{\zeta}_z(\mathbf{e}_{\rm u})=[c_{\zeta_z}^1,\cdots, c_{\zeta_z}^K](\mathbf{C}_{\zeta_z}+\lambda_{\zeta_z}\mathbf{I}_K)^{-1}[\hat{\zeta}_z^1,\cdots, \hat{\zeta}_z^K]^{\rm T},
\end{cases}
\end{equation}
along with the Gaussian-distributed estimation errors.  Since the estimation errors are not related to the following derivations, their details are omitted here. We leverage \eqref{Eq: Regression} to approximate the mapping $\mathbf{e}_{\rm u}\rightarrow(\tilde{\mu}_z,\tilde{\sigma}_z,\tilde{\xi}_z)$.
The details of the elements in \eqref{Eq: Regression} are given as follows:
$c_{\mu_z}^k=\gamma_{\mu_z} e^{ -\frac{\lVert\mathbf{e}_{\rm u}-\bar{\mathbf{e}}_{\rm u}^k\rVert^2}{2l_{\mu_z}}}$,
$c_{\sigma_z}^k=\gamma_{\sigma_z}e^{-\frac{\lVert\mathbf{e}_{\rm u}-\bar{\mathbf{e}}_{\rm u}^k\rVert^2}{2l_{\sigma_z}}}$, and
$c_{\zeta_z}^k=\gamma_{\zeta_z}e^{ -\frac{\lVert\mathbf{e}_{\rm u}-\bar{\mathbf{e}}_{\rm u}^k\rVert^2}{2l_{\zeta_z}}},\forall\,k$. In addition,
$\mathbf{C}_{\mu_z}$, $\mathbf{C}_{\sigma_z}$, and $\mathbf{C}_{\zeta_z}$  are $K\times K$ matrices whose elements in the $k$-th row and $k'$-th column are
$C_{\mu_z}^{kk'}=\gamma_{\mu_z}e^{ -\frac{\lVert\bar{\mathbf{e}}_{\rm u}^k-\bar{\mathbf{e}}_{\rm u}^{k'}\rVert^2}{2l_{\mu_z}}}$,
$C_{\sigma_z}^{kk'}=\gamma_{\sigma_z}e^{-\frac{\lVert\bar{\mathbf{e}}_{\rm u}^k-\bar{\mathbf{e}}_{\rm u}^{k'}\rVert^2}{2l_{\sigma_z}}}$, and
$C_{\zeta_z}^{kk'}=\gamma_{\zeta_z}e^{ -\frac{\lVert\bar{\mathbf{e}}_{\rm u}^k-\bar{\mathbf{e}}_{\rm u}^{k'}\rVert^2}{2l_{\zeta_z}}}$,
respectively. $\mathbf{I}_K$ is the $K\times K$ identity matrix.
Here,
$\gamma_{\mu_z}$, $\gamma_{\sigma_z}$, $\gamma_{\zeta_z}$, $l_{\mu_z}$, $l_{\sigma_z}$, $l_{\zeta_z}$, $\lambda_{\mu_z}$, $\lambda_{\sigma_z}$, and $\lambda_{\zeta_z}$ are the GPR hyperparameters.

Analogous to \eqref{Eq: GEV learning}, the optimal hyperparameters values of the GPR expression \eqref{Eq: Regression} can be acquired by MLE and gradient ascent. The interested readers please refer to \cite[Eq.~(5.8)]{RasWil06} for the detailed mathematical formulae. In addition, there is a built-in function in Matlab to execute GPR and extract the optimal hyperparameter values. For convenience, while running simulations in Section \ref{Sec: Results}, we directly use the Matlab built-in function for GPR instead of programming the iterative gradient ascent algorithm.

\setcounter{equation}{8}
\begin{figure*}
\begin{multline}\label{Eq: GD position}
\mathbf{e}^{q+1}_{\rm u}= 
 \mathbf{e}^{q}_{\rm u}+ \sum\limits_{i\in\mathcal{V}} \frac{-\delta a_i}{(\tilde{\sigma}^q_i\tilde{\zeta}^q_i +\tilde{\mu}^q_i)^2}
  \cdot\Bigg\{\sum_{k=1}^{K}   \frac{-\alpha_{\mu_i}^k \gamma_{\mu_i}(\mathbf{e}^q_{\rm u}-\bar{\mathbf{e}}^k_{\rm u})}{l_{\mu_i}}\cdot  e^{ -\frac{\lVert\mathbf{e}^q_{\rm u}-\bar{\mathbf{e}}_{\rm u}^k\rVert^2}{2l_{\mu_i}}}
   +\left[\sum_{k=1}^{K} \alpha_{\zeta_i}^k \gamma_{\zeta_i} e^{ -\frac{\lVert\mathbf{e}^q_{\rm u}-\bar{\mathbf{e}}_{\rm u}^k\rVert^2}{2l_{\zeta_i}}}\right] 
  \\\cdot \left[\sum_{k=1}^{K} \frac{- \alpha_{\sigma_i}^k\gamma_{\sigma_i} (\mathbf{e}^q_{\rm u}-\bar{\mathbf{e}}^k_{\rm u})}{l_{\sigma_i}}\cdot  e^{-\frac{\lVert\mathbf{e}^q_{\rm u}-\bar{\mathbf{e}}_{\rm u}^k\rVert^2}{2l_{\sigma_i}}}  \right]
+\left[\sum_{k=1}^{K} \alpha_{\sigma_i}^k \gamma_{\sigma_i}e^{-\frac{\lVert\mathbf{e}^q_{\rm u}-\bar{\mathbf{e}}_{\rm u}^k\rVert^2}{2l_{\sigma_i}}}\right]
\cdot\left[\sum_{k=1}^{K} \frac{-\alpha_{\zeta_i}^k\gamma_{\zeta_i} (\mathbf{e}^q_{\rm u}-\bar{\mathbf{e}}^k_{\rm u})}{l_{\zeta_i}}\cdot   e^{-\frac{\lVert\mathbf{e}^q_{\rm u}-\bar{\mathbf{e}}_{\rm u}^k\rVert^2}{2l_{\zeta_i}}} \right] \Bigg\}.
\end{multline}
\noindent\makebox[\linewidth]{\rule{0.84\paperwidth}{0.4pt}}
\end{figure*}
\setcounter{equation}{5}

\section{URLLC-Aware Resolution Selection and UAV Placement}
\label{Sec: Solution}

Now plugging \eqref{Eq: single GEV} into \eqref{Eq: EVT constraint} and incorporating the GPR expression \eqref{Eq: Regression}, we reconsider problem \eqref{Eq: problem EVT} as
\begin{IEEEeqnarray}{cl}
\underset{\mathbf{e}_{\rm u},\mathbf{L}}{\mbox{maximize}}& ~~\sum\limits_{i\in\mathcal{V}} A_{l_i}\notag
\\\mbox{subject to}&~~(1+\tilde{\xi}_i ( a_i/A_{l_i} -\tilde{\mu}_i )/\tilde{\sigma}_i  )^{-\frac{1}{\tilde{\xi}_i}}\leq \epsilon,~\forall\,i\in\mathcal{V},\notag
\\&~~\mbox{\eqref{Eq: problem 3}, \eqref{Eq: problem 4}, and \eqref{Eq: Regression}},\label{Eq: Re-formulated problem}
\end{IEEEeqnarray}
with $a_i=A_1T_{i}^{\rm th} /V$,
which belongs to mix-integer programming problems. Furthermore, as commented at the end of Section \ref{Sec: GEV characterization}, $\tilde{\mu}_i(\mathbf{e}_{\rm u})$, $\tilde{\sigma}_i(\mathbf{e}_{\rm u})$, and $\tilde{\zeta}_i(\mathbf{e}_{\rm u})$ of each VUE $i$ are chosen according to the zone in which the VUE locates. In order to have a tractable solution to problem \eqref{Eq: Re-formulated problem}, we relax $A_{l_i}$ as a continuous variable $B_{i}$ and instead focus on 
\begin{subequations}\label{Eq: Selection and positioning problem}
\begin{IEEEeqnarray}{cl}
\hspace{-1em}\underset{\mathbf{e}_{\rm u},\mathbf{B}}{\mbox{maximize}}&~~\sum\limits_{i\in\mathcal{V}} B_{i}\label{Eq: Selection and positioning problem-1}
\\\hspace{-1em}\mbox{subject to}&~~ (1+\tilde{\xi}_i(a_i/B_{i}-\tilde{\mu}_i )/ \tilde{\sigma}_i )^{-\frac{1}{\tilde{\xi}_i}}\leq \epsilon,~\forall\,i\in\mathcal{V},\label{Eq: Selection and positioning problem-2}
\\\hspace{-1em}&~~B_{i}\geq A_1,~\forall\,i\in\mathcal{V},\label{Eq: Selection and positioning problem-3}
\\\hspace{-1em}&~~\mbox{\eqref{Eq: problem 3} and \eqref{Eq: Regression}},\notag
\end{IEEEeqnarray}
\end{subequations}
with $\mathbf{B}=(B_{i}:i\in\mathcal{V})$. Here, constraint \eqref{Eq: Selection and positioning problem-3} is obtained from constraint \eqref{Eq: problem 4}. The problem-solving procedure is illustrated as follows.

We first fix the UAV's position $\mathbf{e}_{\rm u}$ and obtain the corresponding optimal $\mathbf{B}^{*}(\mathbf{e}_{\rm u})$, which is expressed as the function of $\mathbf{e}_{\rm u}$. Specifically, given a specific position $\mathbf{e}_{\rm u}$ and its corresponding characteristic parameters $(\tilde{\mu}_i,\tilde{\sigma}_i,\tilde{\zeta}_i)$ from GPR, the left-hand-side term of \eqref{Eq: Selection and positioning problem-2} monotonically increases with $B_i$. Hence, by equating \eqref{Eq: Selection and positioning problem-2} and referring to  $\tilde{\zeta}=f(\tilde{\xi})$, we obtain $B^{*}_{i}=\frac{a_i}{ \tilde{\sigma}_i\tilde{\zeta}_i +\tilde{\mu}_i}$ and, accordingly, select the resolution level as $l_i^{*}=\underset{l_i\in\mathcal{L}}{\arg\min}\{B^{*}_i-A_{l_i}\geq 0\}$.
Subsequently, using the obtained $\mathbf{B}^{*}(\mathbf{e}_{\rm u})$, we aim to find the UAV's optimal position $\mathbf{e}^{*}_{\rm u}$ which maximizes \eqref{Eq: Selection and positioning problem-1}. In this regard, we apply $B^{*}_{i}=\frac{a_i}{ \tilde{\sigma}_i\tilde{\zeta}_i +\tilde{\mu}_i}$ to \eqref{Eq: Selection and positioning problem} and have the following non-convex optimization problem:
\begin{subequations}\label{Eq: Positioning problem}
\begin{IEEEeqnarray}{cl}
\underset{\mathbf{e}_{\rm u}}{\mbox{minimize}}&~~\sum\limits_{i\in\mathcal{V}}
 \frac{-a_i}{\tilde{\sigma}_i\tilde{\zeta}_i +\tilde{\mu}_i}\label{Eq: Positioning problem_1}
\\\mbox{subject to}&~~\frac{a_i}{ \tilde{\sigma}_i\tilde{\zeta}_i +\tilde{\mu}_i}\geq A_1,~\forall\,i\in\mathcal{V},\label{Eq: Positioning problem_2}
\\&~~\eqref{Eq: problem 3}\mbox{ and }\eqref{Eq: Regression}.\notag
\end{IEEEeqnarray}
\end{subequations}
To tractably find a solution, we refer to the generalized projected gradient descent method \cite{JaiKar17} in which the iterative gradient descent is taken with respect to the objective function \eqref{Eq: Positioning problem_1} as shown in \eqref{Eq: GD position}. Here, $\delta$ and $q$ are the learning rate and iteration index, respectively. $(\tilde{\mu}_i^q,\tilde{\sigma}_i^q,\tilde{\zeta}_i^q)$ are calculated by applying $\mathbf{e}^q_{\rm u}$ to \eqref{Eq: Regression}. In addition, $\alpha_{\mu_i}^k$, $\alpha_{\sigma_i}^k$, and $\alpha_{\zeta_i}^k$ are the $k$-th elements of the column vectors
$(\mathbf{C}_{\mu_i}+\lambda_{\mu_i}\mathbf{I}_K)^{-1}[\hat{\mu}_i^1,\cdots, \hat{\mu}_i^K]^{\rm T}$, $(\mathbf{C}_{\sigma_i}+\lambda_{\sigma_i}\mathbf{I}_K)^{-1}[\hat{\sigma}_i^1,\cdots, \hat{\sigma}_i^K]^{\rm T}$, and $(\mathbf{C}_{\zeta_i}+\lambda_{\zeta_i}\mathbf{I}_K)^{-1}[\hat{\zeta}_i^1,\cdots, \hat{\zeta}_i^K]^{\rm T}$, respectively.
After $\mathbf{e}^{q+1}_{\rm u}$ is updated in the $q$-th iteration, we check whether the updated value satisfies constraints \eqref{Eq: problem 3} and \eqref{Eq: Positioning problem_2}. If both constraints are satisfied, we execute the update in the next iteration $(q+1)$. Otherwise, we find the point $\hat{\mathbf{e}}^{q+1}_{\rm u}$ in the feasible region which is the closest to $\mathbf{e}^{q+1}_{\rm u}$ and replace the latter with the former for the next iteration update. The steps of the generalized projected gradient descent method are detailed in Algorithm \ref{Alg: Projected GD}.
\begin{algorithm}[t]
 \caption{Generalized Projected Gradient Descent for UAV Placement}
  \begin{algorithmic}[1]
  \State Initialize a feasible point $\mathbf{e}_{\rm u}^{1}$ of \eqref{Eq: Positioning problem} and  $q=1$.
    \Repeat 
      \State Calculate $\mathbf{e}_{\rm u}^{q+1}$ as per \eqref{Eq: GD position}.
      \If{\eqref{Eq: problem 3} and \eqref{Eq: Positioning problem_2} are satisfied}
      \State Update $\mathbf{e}_{\rm u}^{q}\leftarrow\mathbf{e}_{\rm u}^{q+1}$.
       \Else
       \State Find the closest point $\hat{\mathbf{e}}_{\rm u}^{q+1}$, which satisfies \eqref{Eq: problem 3} and \eqref{Eq: Positioning problem_2},  to $\mathbf{e}_{\rm u}^{q+1}$.
       \State Update $\mathbf{e}_{\rm u}^{q}\leftarrow\hat{\mathbf{e}}_{\rm u}^{q+1}$.
       \EndIf
      \State Update $q\leftarrow q+1$.
    \Until{Stopping criteria are satisfied.}
  \end{algorithmic}\label{Alg: Projected GD}
\end{algorithm}

\section{Numerical Results}\label{Sec: Results}

To simulate the UAV placement scheme in the IoV network, we consider a $240\,\mbox{m}\times 240\,\mbox{m}$-area grid road topology with three longitudinal and three latitudinal roads as in Fig.~\ref{Fig: System}. The UAV's height is set as 50\,m. Regarding the air-to-ground channel model, we refer to \cite{NasTuaDuoPoo19} and express the channel gain as $h_i=\frac{\phi\lvert \psi\rvert^2}{(\theta D_i)^2}$
in which $\phi=3.24\times 10^{-4}$ includes the path loss of $-$38.47\,dB and an antenna gain of 2.2846.
The antenna beamwidth is $\theta= \tan^{-1}(24\sqrt{2}/5)$, and $D_i$ denotes the Euclidean distance between the UAV and VUE $i$. Moreover, we consider Rician fading, denoted by $\psi$, with a Rician factor of 12\,dB and $\mathbb{E}[\lvert\psi\rvert^2]=1$. That is, $\psi=\sqrt{\frac{\kappa}{1+\kappa}}e^{j\rho}+\mathcal{C}\mathcal{N}\big(0, \frac{1}{1+\kappa}\big)$ with a uniformly distributed line-of-sight phase $\rho$ and $\kappa=10^{1.2}$.
We consider four video resolution levels $\mathcal{L}=\{240\mbox{p}, 360\mbox{p}, 480\mbox{p}, 720\mbox{p}\}$ with the corresponding image sizes 2.5\,Mbit, 5.5\,Mbit, 9.8\,Mbit, and 22\,Mbit. In addition, the time slot length is 33\,msec. The rest parameters are listed in Table \ref{Table: Parameters}.
For performance comparison, we consider a {\bf fixed} and a {\bf random} UAV placement scheme, which do not require the preliminary information about the maximal delay (for offline GPR model training) and the locations of vehicles (for online UAV placement). In the {\bf fixed} scheme, the UAV statically stays at the center of the geographic area, i.e., $\mathbf{e}_{\rm u}=(0\,\mbox{m},0\,\mbox{m})$. In the {\bf random} scheme, the x and y coordinates are randomly selected between $-$120\,m and 120\,m. In these two baselines, the UAV's height is also fixed at 50\,m.

\begin{table}[t]
\centering
 \caption{Simulation Parameters}
\begin{tabular}{|>{\centering}m{0.6cm}|>{\centering}m{1.1cm}||>{\centering}m{0.6cm}|>{\centering}m{1.6cm}||>{\centering}m{0.6cm}|>{\centering\arraybackslash}m{1.1cm}|}
\hline
 Para. & Value  & Para. & Value &  Para. & Value
\\\hline
  \hline
$V$  &  200  &  $S_0$    &   150 & $P$   & 30\,dBm  
\\  \hline
  $W$  &  50\,MHz  &  $N_0$   &  -174\,dBm/Hz &  $T_{i}^{\rm th}$   &  3\,sec   
\\  \hline
 $\epsilon$  &   $10^{-3}$  &  $d_{\rm th}$  & 20\,m &  $N$   & $3\times 10^{4}$          
 \\  \hline
 $\nu $   &  $5\times 10^{-4}$   & $K$  &  81   &   $\delta $    &  $5\times 10^{-6}$
\\  \hline
 \end{tabular} \label{Table: Parameters}
\end {table}
\begin{figure}[t]
\centering
\includegraphics[width=\columnwidth]{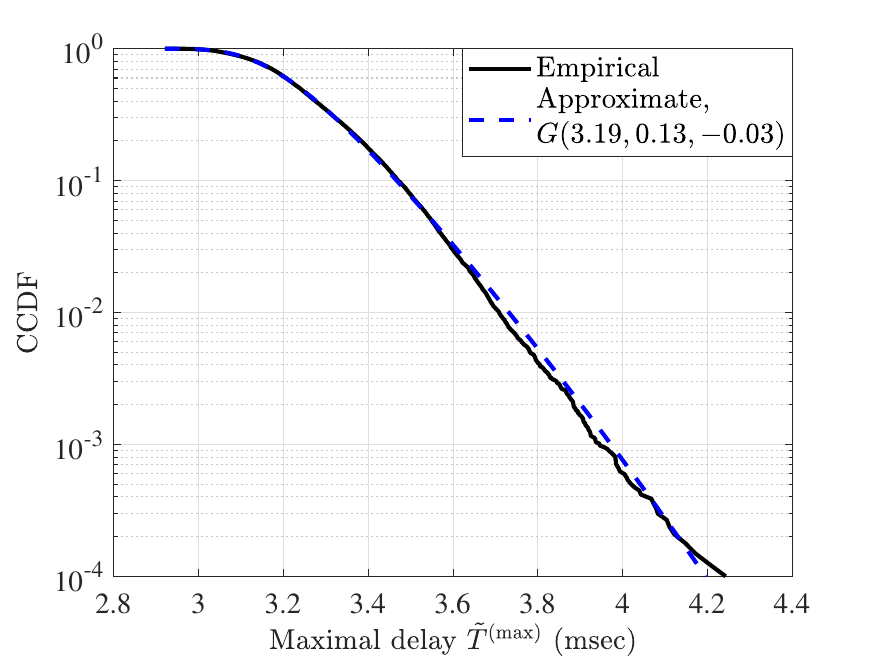}
\caption{Empirical and approximate CCDFs of the maximal delay $\tilde{T}^{(\max)}$ in Zone 1 at $\bar{\mathbf{e}}_{\rm u}=(90\,\mbox{m},60\,\mbox{m})$.}
\label{Fig: EVT} 
\end{figure}
\begin{figure}[t]
\centering
\includegraphics[width=\columnwidth]{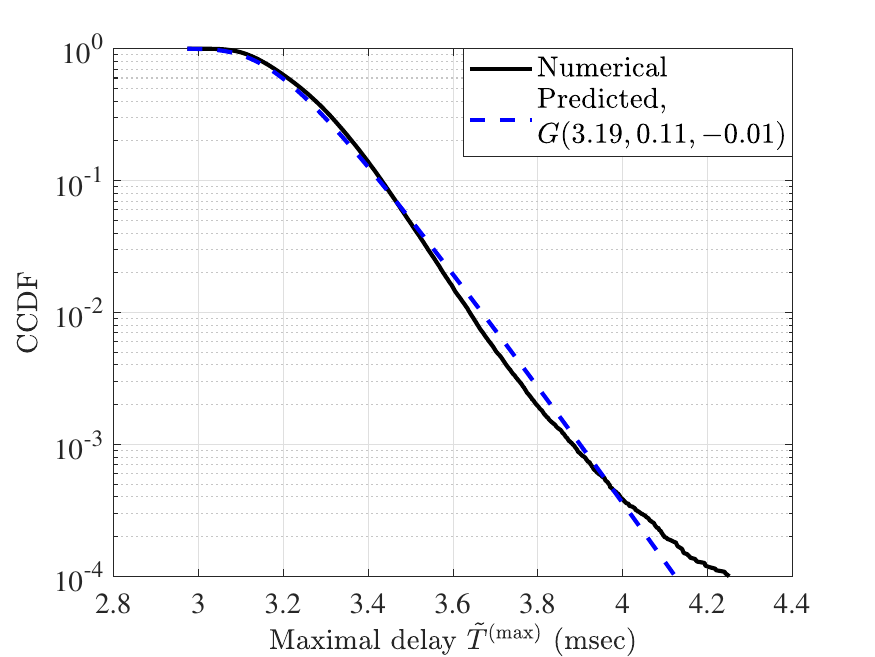}
\caption{Numerical and predicted CCDFs of the maximal delay $\tilde{T}^{(\max)}$ in Zone 6 at $\mathbf{e}_{\rm u}=(45\,\mbox{m}, 45\,\mbox{m})$.}
\label{Fig: GPR estimated}
\end{figure}

We first show the empirical distribution and the approximate GEV distribution (from Theorem \ref{Thm: GEV}) of the maximal delay $\tilde{T}^{(\max)}$ in Fig.~\ref{Fig: EVT}. As the representative, we consider the case in which the UAV locates at $(90\,\mbox{m},60\,\mbox{m})$, and the maximal delays are measured in Zone 1. The similar results can be observed in other cases and zones. The coincidence between the empirical and approximate distributions justifies the usage of EVT for maximal delay characterization.
For GPR, we choose 81 $\bar{\mathbf{e}}_{\rm u}$ points as per the combinations of $\bar{x}_{\rm u}\in\{-120\,\mbox{m},-90\,\mbox{m},\cdots,120\,\mbox{m}\}$ and $\bar{y}_{\rm u}\in\{-120\,\mbox{m},-90\,\mbox{m},\cdots,120\,\mbox{m}\}$. 
After learning the corresponding parameters \eqref{Eq: GEV learning} and acquiring the optimal GPR expression \eqref{Eq: Regression}, we find the predicted parameters for Zone 6 at the testing position $(45\,\mbox{m},45\,\mbox{m})$ as the representative. Fig.~\ref{Fig: GPR estimated} shows the corresponding numerical distribution and the predicted GEV distribution of the maximal delay $\tilde{T}^{(\max)}$. Both curves match well. Similar results can be found in other cases. The result in Fig.~\ref{Fig: GPR estimated} ensures the effectiveness of using GPR to model the position-parameter mapping.

\begin{figure}[t]
\centering
\subfigure[Even distribution of VUEs.]{\includegraphics[width=\columnwidth]{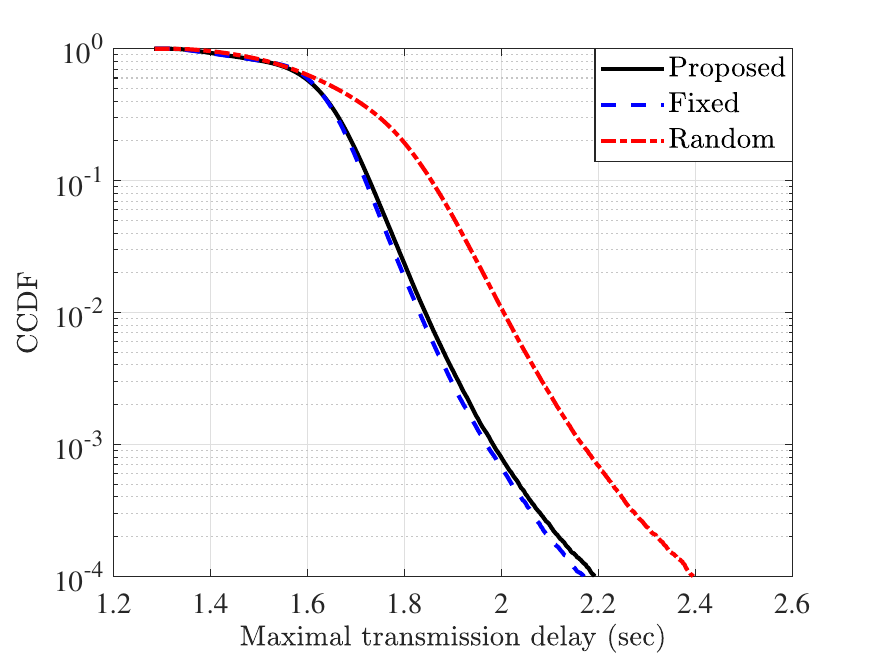}}
\subfigure[Biased distribution of VUEs.]{\includegraphics[width=\columnwidth]{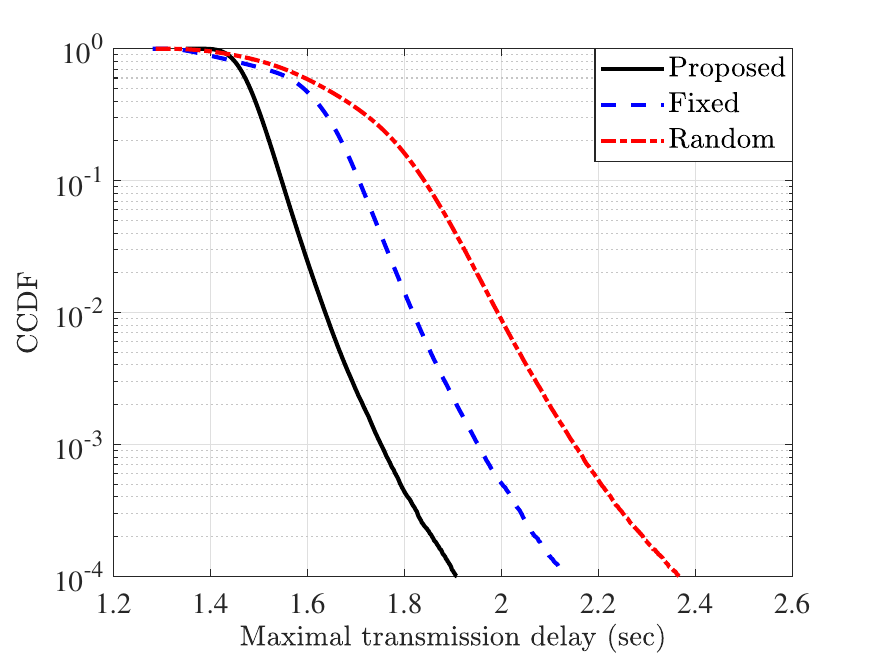}}
\caption{CCDFs of the maximal transmission delay in different UAV placement schemes.}
	\label{Fig: Proposed vs baselines}
\end{figure}

Subsequently, we examine the performance of the maximal transmission delay in our proposed UAV placement approach and compare it with the {\bf fixed} and {\bf random} schemes in Fig.~\ref{Fig: Proposed vs baselines}. To ensure the comparison fairness, the resolution levels in the two baselines are selected as the average resolution in the proposed approach. When the VUEs move evenly in the geographic area, it can be straightforwardly expected that the optimal UAV position is at the area's center. In other words, the {\bf fixed} scheme gives the optimal performance. From Fig.~\ref{Fig: Proposed vs baselines}(a), we can see that the proposed UAV placement approach almost achieves the optimal performance. In the biased distribution in which the VUEs mainly move in both Zones 1 and 6, the proposed approach outperforms both baselines. Quantitatively, the 999-th 1000-quantile of the maximal transmission delay in the proposed approach is $10\%$ and $19\%$ lower than the {\bf fixed} and {\bf random} schemes, respectively.

\begin{figure}[t] 
\centering
\includegraphics[width=\columnwidth]{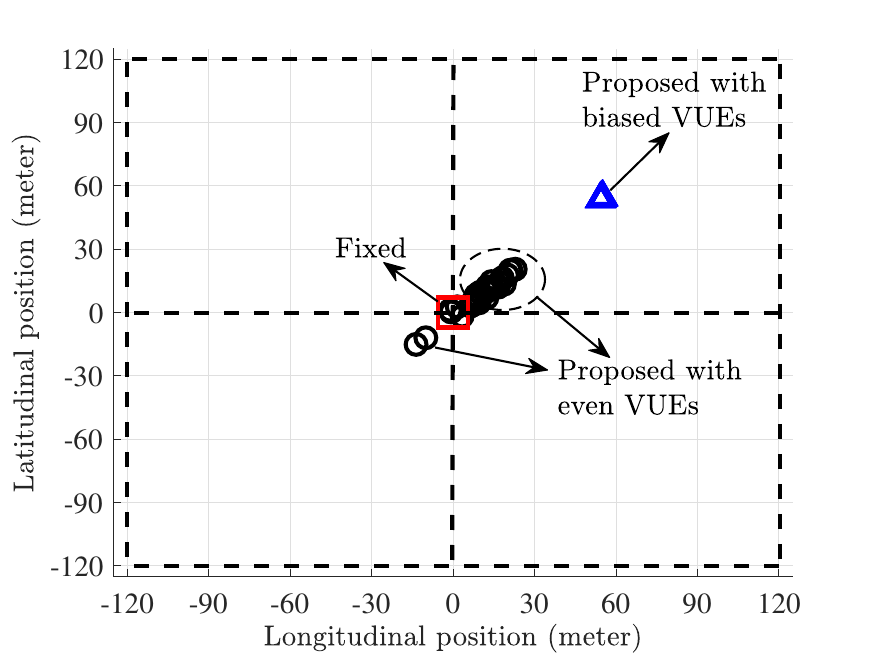}
\caption{The UAV's placed positions (projected onto the xy-plane) in the proposed scheme. Dash lines represent the roads.}
\label{Fig: UAV placement}
\end{figure}

To give further insights of the results in Fig.~\ref{Fig: Proposed vs baselines}, let us show the UAV's placed positions of the proposed approach in Fig.~\ref{Fig: UAV placement}. When the VUEs are evenly distributed, the UAV is placed around the center of the geographic area. Since the placed positions of our approach and the {\bf fixed} scheme are close, the delay distribution curves  in Fig.~\ref{Fig: Proposed vs baselines}(a) almost coincide. In the biased case, the UAV in our approach is adaptively placed within Zones 1 and 6, whereas the UAV in the {\bf fixed} scheme is kept at the center. This adaptability of the proposed approach gives the delay performance enhancement in Fig.~\ref{Fig: Proposed vs baselines}(b).

\section{Conclusion}
\label{Sec: Conclusion}
We have studied the UAV placement problem in the IoV network, which is cast as video resolution maximization subject to the URLLC requirement for the maximal transmission delay. By leveraging EVT and GPR, we have come up with a way to approximate the mapping function between the UAV's position and the maximal delay performance. Then we have proposed a proactive approach for resolution selection and UAV placement. In numerical results, we have verified the effectiveness of jointly using EVT and GPR for maximal delay characterization. Moreover, we have showed that our proposed UAV placement approach nearly achieves the optimal performance of the maximal transmission delay in the even VUE distribution. In the biased VUE distribution, our approach has achieved the 10\% and 19\% reductions of the 999-th 1000-quantile over the {\bf fixed} and {\bf random} placement schemes, respectively. Finally, we have shown that the proposed approach can adaptively place the UAV.

\bibliographystyle{IEEEtran}
\bibliography{Ref_IoV}

\begin{thebibliography}{10}
\providecommand{\url}[1]{#1}
\csname url@samestyle\endcsname
\providecommand{\newblock}{\relax}
\providecommand{\bibinfo}[2]{#2}
\providecommand{\BIBentrySTDinterwordspacing}{\spaceskip=0pt\relax}
\providecommand{\BIBentryALTinterwordstretchfactor}{4}
\providecommand{\BIBentryALTinterwordspacing}{\spaceskip=\fontdimen2\font plus
\BIBentryALTinterwordstretchfactor\fontdimen3\font minus
  \fontdimen4\font\relax}
\providecommand{\BIBforeignlanguage}[2]{{%
\expandafter\ifx\csname l@#1\endcsname\relax
\typeout{** WARNING: IEEEtran.bst: No hyphenation pattern has been}%
\typeout{** loaded for the language `#1'. Using the pattern for}%
\typeout{** the default language instead.}%
\else
\language=\csname l@#1\endcsname
\fi
#2}}
\providecommand{\BIBdecl}{\relax}
\BIBdecl

\bibitem{XuZhoCheLyuShiCheShe18}
W.~Xu, H.~Zhou, N.~Cheng, F.~Lyu, W.~Shi, J.~Chen, and X.~Shen, ``Internet of
  {Vehicles} in big data era,'' \emph{IEEE/CAA J. Automatica Sinica}, vol.~5,
  no.~1, pp. 19--35, Jan. 2018.

\bibitem{ShiZhoLiXuZhaShe18}
W.~Shi, H.~Zhou, J.~Li, W.~Xu, N.~Zhang, and X.~Shen, ``Drone assisted
  vehicular networks: Architecture, challenges and opportunities,'' \emph{IEEE
  Netw.}, vol.~32, no.~3, pp. 130--137, May/Jun. 2018.

\bibitem{LiuZhuDenGuaWanLuoLinZha19}
Y.~Liu, C.~Zhu, X.~Deng, P.~Guan, Z.~Wan, J.~Luo, E.~Liu, and H.~Zhang,
  ``{UAV}-aided urban target tracking system based on edge computing,''
  \emph{CoRR}, vol. abs/1902.00837, pp. 1--6, Feb. 2019.

\bibitem{LiaMaHuaWan23}
Z.~Liao, Y.~Ma, J.~Huang, and J.~Wang, ``Energy-aware {3D}-deployment of {UAV}
  for {IoV} with highway interchange,'' \emph{IEEE Trans. Commun.}, vol.~71,
  no.~3, pp. 1536--1548, Mar. 2023.

\bibitem{LiuLaiLinLeu22}
X.~Liu, B.~Lai, B.~Lin, and V.~C.~M. Leung, ``Joint communication and
  trajectory optimization for multi-{UAV} enabled mobile {Internet of
  Vehicles},'' \emph{IEEE Trans. Intell. Transp. Syst.}, vol.~23, no.~9, pp.
  15\,354--15\,366, Sep. 2022.

\bibitem{CaiFenHeXuZhaXie20}
R.~Cai, Y.~Feng, D.~He, Y.~Xu, Y.~Zhang, and W.~Xie, ``Trajectory optimization
  for large-scale {UAV}-assisted {RSU}s in {V2I} communication,'' in
  \emph{Proc. IEEE 92nd Veh. Technol. Conf.}, Nov. 2020, pp. 1--6.

\bibitem{DemTokEki20}
U.~Demir, C.~Toker, and {\"O}.~Ekici, ``Energy-efficient deployment of {UAV} in
  {V2X} network considering latency and backhaul issues,'' in \emph{Proc. IEEE
  Int. Black Sea Conf. Commun. Netw.}, May 2020, pp. 1--6.

\bibitem{YanLiaZha21}
J.~Yang, T.~Liang, and T.~Zhang, ``Deployment optimization in {UAV} aided
  vehicle localization,'' in \emph{Proc. IEEE 93rd Veh. Technol. Conf.}, Apr.
  2021, pp. 1--6.

\bibitem{LiuBen18}
C.-F. Liu and M.~Bennis, ``Ultra-reliable and low-latency vehicular
  transmission: An extreme value theory approach,'' \emph{IEEE Commun. Lett.},
  vol.~22, no.~6, pp. 1292--1295, Jun. 2018.

\bibitem{BatLiuBenSurHon20}
S.~Batewela, C.-F. Liu, M.~Bennis, H.~A. Suraweera, and C.~S. Hong,
  ``Risk-sensitive task fetching and offloading for vehicular edge computing,''
  \emph{IEEE Commun. Lett.}, vol.~24, no.~3, pp. 617--621, Mar. 2020.

\bibitem{YuTakKaiSak21}
T.~Yu, Y.~Takaku, Y.~Kaieda, and K.~Sakaguchi, ``Design and {PoC}
  implementation of {mmWave}-based offloading-enabled {UAV} surveillance
  system,'' \emph{IEEE Open J. Veh. Technol.}, vol.~2, pp. 436--447, 2021.

\bibitem{Coles01}
S.~Coles, \emph{An Introduction to Statistical Modeling of Extreme
  Values}.\hskip 1em plus 0.5em minus 0.4em\relax London, U.K.: Springer, 2001.

\bibitem{AlsLinNiyTan14}
M.~A. Alsheikh, S.~Lin, D.~Niyato, and H.-P. Tan, ``Machine learning in
  wireless sensor networks: Algorithms, strategies, and applications,''
  \emph{IEEE Commun. Surveys Tuts.}, vol.~16, no.~4, pp. 1996--2018, 4th Quart.
  2014.

\bibitem{RasWil06}
C.~E. Rasmussen and C.~K.~I. Williams, \emph{Gaussian Processes for Machine
  Learning}.\hskip 1em plus 0.5em minus 0.4em\relax Cambridge, MA, USA: MIT
  Press, 2006.

\bibitem{JaiKar17}
P.~Jain and P.~Kar, ``Non-convex optimization for machine learning,''
  \emph{Found. Trends Mach. Learn.}, vol.~10, no. 3--4, pp. 142--363, 2017.

\bibitem{NasTuaDuoPoo19}
A.~A. Nasir, H.~D. Tuan, T.~Q. Duong, and H.~V. Poor, ``{UAV}-enabled
  communication using {NOMA},'' \emph{IEEE Trans. Commun.}, vol.~67, no.~7, pp.
  5126--5138, Jul. 2019.

\end{thebibliography}

\end{document}